\documentclass{emulateapj}
\usepackage{apjfonts}
\usepackage{natbib}
\usepackage{amsmath}
\usepackage{graphicx}
\def \ni {\noindent}
\def\lsim{\mathrel{\rlap{\lower4pt\hbox{\hskip1pt$\sim$}}
    \raise1pt\hbox{$<$}}}                
\def\gsim{\mathrel{\rlap{\lower4pt\hbox{\hskip1pt$\sim$}}
    \raise1pt\hbox{$>$}}}                
\def\fml{$M_*/L_V$}
\def\fmls{$M_*/L_V$ }

\def\sersics{S\'{e}rsic }

\def\apjl{ApJL}
\def\aj{AJ}
\def\mnras{MNRAS}
\def\apj{ApJ}
\def\araa{ARA\&A}

\def\aap{A\&A}

\defcitealias{nav96}{NFW}
\defcitealias{jar13}{J13}

\slugcomment{{\sc Accepted to ApJL:} 16 August 2013} 
\shorttitle{Non-parametric Schwarzschild Models of the dSphs}

\begin{document}
\title{Variations in a Universal Dark Matter Profile for Dwarf Spheroidals}



\author{John R. Jardel and Karl Gebhardt}

\affil{Department of Astronomy, University of Texas at
Austin, 1 University Station C1400, Austin, TX 78712;\\
jardel@astro.as.utexas.edu}

\begin{abstract}

Using a newly-developed modeling technique, we present orbit-based dynamical 
models of the Carina, Draco, Fornax, Sculptor, and Sextans dwarf spheroidal 
(dSph) galaxies.  These models calculate the dark matter profiles 
non-parametrically without requiring any assumptions to be made about 
their profile shapes.  By lifting this restriction we discover a host
of dark matter profiles in the dSphs that are different from the typical
profiles suggested by both theorists and observers.  However, when we scale
these profiles appropriately and plot them on a common axis they appear
to follow an approximate $r^{-1}$ power law with considerable scatter.  

\end{abstract}


\section{Introduction}

It is a well known fact that cosmological simulations containing only 
collisionless dark matter produce halos that share a universal density
profile $\rho_{\mathrm{DM}}(r)$  \citep{nav96,spr08,nav10}. 
At first, this universal profile was characterized by the double power-law
Navarro-Frenk-White (NFW) profile \citep{nav96}
with inner logarithmic slope $\alpha \equiv - d\log \rho_{\mathrm{DM}}/d\log r = 1$.
Modern dark matter-only simulations with increasingly better resolution seem 
to produce profiles that, in analogy to the \sersics function \citep{ser68},
transition smoothly from $\alpha=3$ in the outer regions 
to $\alpha \sim 1$ near the center \citep{mer05,gao08,nav10}.
The exact form of $\rho_{\mathrm{DM}}(r)$ is still debated by theorists, 
but most agree that the inner slope is nonzero.
Such profiles are called ``cuspy'' since $\rho_{\mathrm{DM}}$ increases as
$r \to 0$.  In contrast, observers modeling low-mass galaxies with
stellar and gas dynamics often find $\alpha=0$ ``cores'' in the inner profiles
\citep{bur95,per96,bor01,deb01,sim05}.  This disagreement between theory
and observations has become known as the core/cusp debate.  

We must remember, however, that real galaxies are the products of their 
unique formation histories, 
and complex baryonic processes can re-shape dark matter
profiles in different ways.  Whether originating 
from adiabatic compression \citep{blu86}, supernovae winds \citep{nav96b}, or
ram-pressure stripping \citep{arr12}, baryonic feedback has been shown to affect
the dark matter profiles of galaxies by perturbing their baryons in a 
highly non-linear 
way. Since these processes differ on a galaxy-by-galaxy basis, one should not 
expect to observe a universal dark matter profile at $z=0$.  Furthermore, 
given the 
number of different ways baryonic feedback can occur, we should not expect it to
produce only cored or NFW-like profiles.

Unfortunately, it is difficult to explore the possible range of profile shapes
since
to construct a dynamical model one generally needs to adopt a parameterization 
for $\rho_{\mathrm{DM}}(r)$.  This is not ideal as one is forced to assume the 
very thing they are hoping to measure.  Clearly methods that can measure 
$\rho_{\mathrm{DM}}(r)$ non-parametrically are
advantageous.  Non-parametric determination of the dark matter profile 
avoids biasing results by assuming an incorrect parameterization and it
also allows more general profile types to be discovered.

To test the universal profile assumption, 
we apply the technique of non-parametric Schwarzschild modeling
to determine $\rho_{\mathrm{DM}}(r)$
in five of the brightest dwarf spheroidal (dSph) galaxies
that orbit the Milky Way as satellites.  These galaxies have excellent
kinematics available \citep{wal09} and have been demonstrated to be good
targets for this type of modeling \citep{jar13}.  The dSphs as a 
population are some of the most
dark matter-dominated galaxies ever observed \citep{mat98,sim07} and
as such are unique test sites for theories of galaxy formation at low mass
scales.

Past studies using Jeans models
have had difficulty robustly measuring $\rho_{\mathrm{DM}}(r)$ in the dSphs 
\citep{wal09b}
largely due to the degeneracy between mass and velocity anisotropy inherent to 
these models.
In addition to being fully non-parametric, our models break 
the degeneracy between
mass and velocity anisotropy the same way traditional Schwarzschild models
accomplish this \citep{geb00,rix97,vdm98,val04,vdb08}.  
In this letter we apply the most
general models to a widely-studied group of galaxies in order to
measure their dark mater density profiles and test the universal
profile hypothesis.

\section{Data}

Our models use the publicly available kinematics data from \citet{wal09} for
Carina, Fornax, Sculptor and Sextans.  These data are individual radial
velocities for member stars with repeat observations weighted and averaged.
\citet{wal09} assign each star a membership probability $P$ based on its
position, velocity, and a proxy for its metallicity.  Our analysis only includes
stars for which $P > 0.95$.  Whenever a§ galaxy has high-quality
\emph{Hubble Space Telescope} measurements of its proper motion available
(Carina and Fornax; \citealt{pia03,pia07}) we correct for
the effects of perspective rotation following Appendix A of \citet{wal08}.

As described in \citet{jar12}, stars are placed on a meridional grid according
to their positions and folded over the major and minor axes.  To preserve any
possible rotation, we switch the sign of the velocity whenever a star is flipped
about the minor axis.  We then group the stars into spatial bins by dividing
the grid into a series of annular bins containing roughly 50-70 stars per bin.
Fornax and Sculptor have a larger number of stars with measured 
velocities, and to exploit
this we subdivide the annular bins into two to three angular bins in
analogy to spokes on a wheel.  Table \ref{sum} presents a summary
of the data we use for the dSphs.

For each spatial bin of stars, we reconstruct the full line-of-sight velocity
distribution (LOSVD) from the discrete radial velocities observed.  This
procedure uses an adaptive kernel density estimator \citep{sil86} and is
described in more detail in \citet{jar13}.  Uncertainties in the LOSVDs are
determined through bootstrap resamplings of the data.  We divide each LOSVD 
into 15 velocity bins which serve as the observational constraint for our
models. 

Also necessary for the models is the galaxy's three-dimensional luminosity
density profile $\nu(r)$.  To obtain this, we start with the projected
number density profile of stars $\Sigma_*(R)$.  For Carina and Sculptor we 
take $\Sigma_*(R)$ from \citet{wal03}, opting to
use their fitted King profile for Carina and the actual profile for Sculptor 
with no fit performed.  We also use a King profile to describe $\Sigma_*(R)$
in Sextans, with the parameters taken from \citet{irw95}.  In Fornax we 
use the full profile reported in \citet{col05}.
We then convert $\Sigma_*(R)$ to a surface brightness 
profile $\mu(R)$ by adding an arbitrary zero-point shift, in log space, and
adjusting the shift until the integrated $\mu(R)$ returns a luminosity 
consistent with the value listed in \citet{mat98}.  

Next we deproject 
$\mu(R)$ via Abel inversion through the manner described in \citet{geb96}.
For simplicity in the deprojection and subsequent modeling, we assume that
each galaxy is viewed edge-on.  
For a thorough discussion on how uncertainties in viewing angle
and geometry propagate through our models we refer the reader to 
\citet{tho07b}.  Our models are axisymmetric, so we 
use the stellar ellipticity to determine $\nu$ away from the major
axis.  

\begin{deluxetable}{llllll}
\tablecaption{Properties of the Dwarf Spheroidals}
\tablewidth{0pt}
\tablehead{
  \colhead{Galaxy} & \colhead{Distance (kpc)} & 
  \colhead{$N_{\mathrm{stars}}$} & \colhead{$N_{\mathrm{LOSVD}}$} & 
  \colhead{Ellipticity} & \colhead{$R_{\mathrm{trunc}}$}}
\startdata
Carina & $104^a$ & $702^f$ & 14 & $0.33^e$ & 4.2 \\
Draco & $71^b$ & $170^{gh}$ & 8 & $0.29^e$ & 3.1\\
Fornax & $136^a$ & $2409^f$ & 36 & $0.30^e$ & 13.5\\
Sculptor & $85^c$ & $1266^f$ & 24 & $0.32^e$ & 5.1\\
Sextans & $85^d$ & $388^f$ & 8 & $0.35^e$ & 5.1
\enddata
\label{sum}
\tablecomments{Summary of the data we use for our study of the dSphs.  We list
the distances to the dSphs we have assumed, the number of member stars
with radial velocity measurements $N_{\mathrm{stars}}$, and 
the number of LOSVDs these measurements are divided into $N_\mathrm{LOSVD}$.  
We assume the the dark matter halo has the same ellipticity as the value 
listed for the stellar component. We also list the truncation
radius $R_{\mathrm{trunc}}$ used in our analysis. References:  
$^a$\citet{tam08}, $^b$\citet{ode01},
$^c$\citet{pie08}, $^d$\citet{lee09}, $^e$\citet{irw95}, 
$^f$\citet{wal09}, $^g$\citet{kle02}, $^h$\citet{jar13}.   }

\end{deluxetable}


\section{Models}

The non-parametric modeling technique we use is described in full detail in
\citet{jar13}.  It is based on the Schwarzschild modeling code of 
\citet{geb00} updated by \citet{tho04,tho05} and described in
\citet{sio09}.  
We have tested our models by using kinematics
generated from a Draco-sized mock dSph embedded in a larger dark matter halo with
either a cored or NFW-like cuspy profile.  In both cases we
are able to accurately recover the density profile from which the mock
kinematics were drawn.

The fundamental principle behind Schwarzschild modeling, that of 
orbit superposition, was first introduced by \citet{sch79}.  
The Schwarzschild code that is the backbone of our non-parametric technique
has been thoroughly tested using artificial data.  It has been shown to 
accurately recover the mass profile and orbit structure of simple isotropic 
rotators \citep{tho05}, $N$-body merger remnants \citep{tho07b}, and a mock 
galaxy containing a supermassive black hole \citep{sio09}.  The
general Schwarzschild technique has also been tested with artificial
data representing the binned individual velocities typically used as
input for studying the dSphs \citep{bre12}.

This method
works by assuming a trial potential for the galaxy under study and 
determining all stellar orbits that are possible in that potential.  
Our orbit sampling scheme is described in detail in \citet{tho04}.
The orbits are then assigned weights according to how well they match the
LOSVDs and a $\chi^2$ value is determined, subject to a constraint of
maximum entropy \citep{sio09}.  If $\chi^2$ is low, the orbits
are a good fit to the kinematics and the trial potential is considered to be a
good estimate for the real potential.  If $\chi^2$ is large, the trial potential
does not support orbits that can match the kinematics and a new potential
is generated.  Each model is required to match $\nu(r)$ as well to machine
precision.

We construct the many trial potentials by solving Poisson's equation for a
specified total density profile
$\rho(r)$ along the major axis.  We assume the total mass distribution has the
same ellipticity as the stellar component and use this adopted ellipticity to
define $\rho(r,\theta)$ away from the major axis. 

Rather than paramaterizing
$\rho(r)$ with an unknown function and sampling its parameters, we take an
altogether different approach (detailed in \citealt{jar13}).  To 
describe $\rho(r)$ we divide the profile into 5 radial points $r_i$, 
equally spaced in $\log r$.  A trial $\rho(r)$ is then represented by the 
density $\rho_i$ at each point.  In this way, the $\rho_i$ themselves are the
parameters that we adjust when picking trial potentials.  To sample this 
parameter space, we employ a similar iterative refinement scheme as discussed in
\citet{jar13}.  We also impose the same constraint that each profile 
must be non-increasing as a function of radius.

Since the dSphs orbit within the Milky Way's halo, the possibility exists that
they are being, or have been, tidally stripped.  In constructing our trial
potential, we account for this by leaving the slope of $\rho(r)$ outside of
our model grid a free parameter $\alpha_{\infty}$.  Each model profile $\rho(r)$
is run with $\alpha_{\infty} \in \{2, 3, 4\}$.  In this way, we treat 
$\alpha_{\infty}$ as a nuisance parameter and marginalize 
over it for the rest of our discussion.  We also truncate the dSphs at the 
radius $R_{\mathrm{trunc}}$ defined by the Jacobi radius given the mass of the 
dSph, its Galactocentric distance, and the mass of the Milky Way (assumed to be
$M_{\mathrm{MW}}=3 \times 10^{12} \, M_{\odot}$ and represented by an isothermal 
sphere).  We list values for $R_{\mathrm{trunc}}$ in Table \ref{sum}.

\begin{figure*}[t]
  \centering
  \includegraphics[width=15cm]{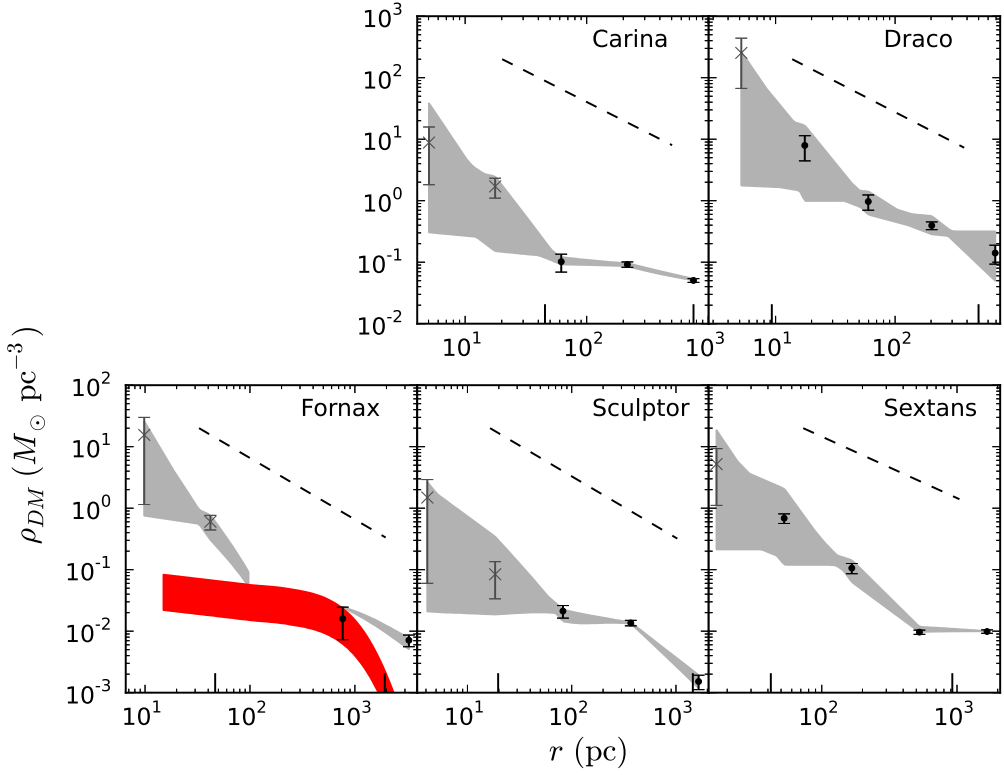}
  \caption{Dark matter density profiles for each dSph modeled as well as 
    Draco from 
    \citet{jar13}.  Points with error bars show the $\Delta \chi^2=1$ 
    uncertainties in the $\rho_i$.  These points are gray X's when they lie
    interior to the our kinematics and black dots otherwise.  Vertical 
    black tick marks on the x-axes show the radial extent of this range.  The
    gray shaded regions show the joint confidence range of the entire profile
    at the level of $\Delta \chi^2=5.84$ and interpolated between the 
    $\rho_i$.  Each panel plots a generic $r^{-1}$ NFW-like profile as a dashed
    line.  In Fornax we also plot the stellar density profile $\rho_*(r)$ 
    with $1\sigma$ uncertainties in red.  
    \label{all}}
\end{figure*}

\subsection{Stellar Density}

After running a large number of models for each galaxy, we have a non-parametric
measurement of the \emph{total} density profile $\rho(r)$.  In order to 
obtain the dark matter density profile we must subtract the stellar density 
$\rho_*(r)$.  This requires knowledge of the stellar mass-to-light ratio
$M_*/L_V$ since $\rho_*(r) = M_*/L_V \times \nu(r)$, assuming that variations
in \fmls with radius are unimportant.

To estimate \fmls we use photometrically derived determinations of each 
galaxy's stellar age $t_{\mathrm{age}}$
and metallicity [Fe/H] \citep{lia11}.  The simple stellar population models of 
\citet{mar05} then yield an estimate on \fmls given these two quantities and
the assumption of either a Salpeter or Kroupa initial mass function (IMF).  We
characterize the uncertainty in $\rho_*(r)$ by the spread in values of
\fmls that result from a choice in IMF and the uncertainties in 
$t_{\mathrm{age}}$ and [Fe/H].  For our analysis of $\rho_{\mathrm{DM}}(r)$,
we add in quadrature the uncertainties on $\rho(r)$ from the models with
those on $\rho_*$ due to \fml.

In all but one of the dSphs (Fornax), $\rho(r) \gg \rho_*(r)$ making the 
determination of $\rho_*(r)$ relatively unimportant.  In Fornax, however, the 
relatively large uncertainties on $\rho_*(r)$ make $\rho(r) - \rho_*(r)$ 
a negative quantity in some cases.  This is clearly unphysical as it represents
a negative dark matter density.  To better study Fornax and other relatively 
baryon-dominated galaxies, a more accurate determination of \fmls is
required.

\subsection{$\chi^2$ Analysis}

We evaluate the goodness of fit of each model with $\chi^2$ as calculated by 

\begin{equation}
\chi^2 = \sum_{i=1}^{N_{\mathrm{LOSVD}}} \sum_{j=1}^{N_{\mathrm{vel}}=15}
\left ( \frac{\ell_{ij}^{\mathrm{obs}} - \ell_{ij}^{\mathrm{mod}}}
{\sigma_{ij}} \right )^2 ,
\label{chi2eq}
\end{equation}

\ni
where the sums are computed over the $N_{\mathrm{vel}}=15$ velocity bins for 
all of the  LOSVDs in each galaxy.  The $\ell_{ij}$ correspond to 
the value in the jth velocity bin of the ith LOSVD.  The uncertainty in 
$\ell^{\mathrm{obs}}_{ij}$ is $\sigma_{ij}$.  

We identify the best-fitting model as that which has the lowest value of 
the (unreduced) $\chi^2=\chi^2_{\mathrm{min}}$. A na\"{i}ve calculation of the 
reduced 
$\chi^2_{\nu}= \chi^2_{\mathrm{min}}/(N_{\mathrm{vel}} \times N_{\mathrm{LOSVD}})$ 
often yields values much less than unity due to correlation between velocity 
bins caused by our kernel density estimator.  We instead test for the 
overall goodness of fit of our best model by computing $\chi^2_{\nu,GH}$:
the reduced $\chi^2$ with respect to a Gauss-Hermite parameterization
of our best-fitting LOSVDs.  We find $\chi^2_{\nu,GH}$ ranges from  $0.3-0.9$ for
the four dSphs modeled here.  These values are consistent with past results
\citep{geb03,geb09,jar13} and have been demonstrated to lead to accurate 
recovery 
of the mass profiles of mock galaxies \citep{tho05}.  We therefore scale
our model-computed unredoced $\chi^2$ values by a factor equal to 
$\chi^2_{\nu} / \chi^2_{\nu,GH}$ in order to bring our reduced $\chi^2_{\nu}$ 
nearer to $\chi^2_{\nu,GH}$.

We present our dark matter profiles at two different levels of confidence.  
When 
specifying the dark matter density at a single point, we marginalize over 
all other parameters using the sliding boxcar technique described 
in \citet{jar13} to interpolate
$\chi^2$.  The 1$\sigma$ confidence interval thus corresponds to a limit
of $\Delta \chi^2=1$ (for one degree of freedom) above $\chi^2_{\mathrm{min}}$.  
When referring to the joint 1$\sigma$ confidence interval of the entire profile,
we instead include limits derived from all models within $\Delta \chi^2=5.84$
of $\chi^2_{\mathrm{min}}$ (for 5 degrees of freedom).

\clearpage

\section{Results}

\begin{figure*}
  \centering
  \includegraphics[width=12cm]{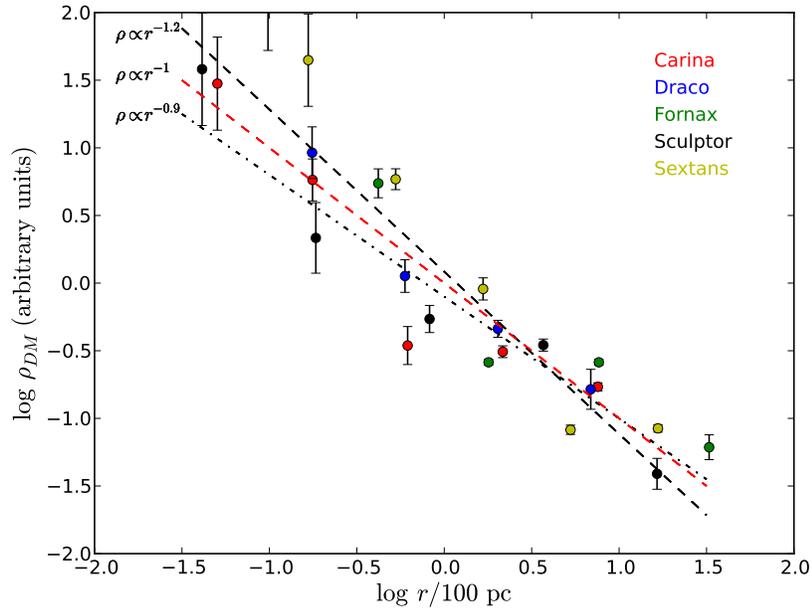}
  \caption{Combined dark matter density profiles of all the dSphs 
    plotted on the same
    axes.  Each galaxy's profile is plotted with the same colored points.  
    Uncertainties on these points are the $\Delta \chi^2=1$ uncertainties from
    Figure \ref{all}.  We plot the derived best-fit line with slope 
    $\alpha = 1.2 \pm 0.5$ as a dashed line as well as the NFW profile with
    $\alpha = 1.0$ as a red dashed line.  A fit excluding the points where
    we have no kinematics available is shown as a dotted line. 
    The individual profiles have been scaled to a common height.
    \label{comb}}
\end{figure*}

We present the non-parametrically determined dark matter profiles in 
Figure \ref{all}.
In addition to the new results for Carina, Fornax, Sculptor, and Sextans, we 
include the result from \citet{jar13} for Draco.  Each panel in Figure 
\ref{all} contains a dashed line with $\rho_{\mathrm{DM}} \propto r^{-1}$ to 
show the
generic shape of the NFW profile.  
The points with error bars in Figure \ref{all} are the marginalized dark matter
density determined from $\Delta \chi^2=1$ at the $r_i$ where the
total density is being varied from model to model.  The gray points labeled with
X's are located interior to the radial range over which stellar kinematics 
are available.  We denote this range for each galaxy with vertical tick marks
on the x-axis.
The joint confidence band (shaded region) interpolates between the $r_i$ 
by taking the maximum and minimum value for $\rho_{\mathrm{DM}}$ at each radius 
for every model within $\Delta \chi^2=5.84$ of $\chi^2_{\mathrm{min}}$.

Given the freedom to choose a dark matter profile of any shape,
it is immediately apparent that our models have chosen a 
variety of shapes for the dSphs.  Draco appears the most similar
to the NFW profile while Sculptor most closely resembles a broken powerlaw
that becomes shallower towards its center.
The other galaxies host profiles that resemble neither cores nor cusps: 
Carina's profile appears flat where
we have kinematics but then displays a possible up-bending 
inside of this region.  Sextans has a steeper slope than the NFW profile until
its outermost point where it suddenly becomes flat.  These sharp differences
among dSph dark matter profiles demonstrate the variety of profile shapes in 
the Local Group.


Unfortunately, due to a lack of central stellar velocities in the \citet{wal09} 
data, the central profiles of the dSphs we model become increasingly 
uncertain there.  This is evidenced by the larger error bars on our gray
points in Figure \ref{all} where we have no kinematics coverage.  However,
we do have some constraint from projection effects and radial orbits in our
models that have apocenters at radii where we do have data.  

\newpage

\subsection{Fornax}

Fornax is an especially difficult case for non-parametric modeling
because, compared to the
other dSphs, it is relatively baryon-dominated.  Our imprecise determination of 
\fmls in Fornax causes $\rho_*(r)$ to be greater than the total modeled density
at some radii, making $\rho_{\mathrm{DM}}(r)$ negative.  In our analysis of 
Fornax, we do not plot the radial range over which this occurs as it is
unphysical.  Instead in Figure \ref{all} we over-plot the stellar density in
red to illustrate why the subtraction is difficult in Fornax.  In all other 
panels $\rho_*(r) \ll \rho_{\mathrm{DM}}(r)$ and is not plotted.

There is strong evidence from multiple studies using independent methods that
suggests Fornax has a dark matter profile that is not cuspy like the 
NFW profile.  
\citep{goe06,wal11,jar12}.  Each of these studies only contrasts between cored
and cuspy profiles or uses a single slope to characterize the profile. 
It is therefore interesting to
explore the non-parametric result we obtain.  Even though we cannot determine
$\rho_{DM}$ where the stellar density is greater than the total density,
we can still place an upper limit on $\rho_{DM}$ such that it must not be 
greater than $\rho_*$ or the red band in Figure \ref{all}.  Given this 
constraint, we can see that the outer profile of Fornax is flat, while the 
inner portion rises more steeply than $r^{-1}$.  Past dynamical studies of 
Fornax only compared generic cored and NFW profiles and did not test this
up-bending profile, therefore it is difficult to compare to their results.

\subsection{A Common Halo?}

Despite the differences in the individual profiles of the dSphs, when we plot
them on the same axes they appear to follow a combined
$r^{-1}$ profile with scatter.  We plot this combined profile in Figure 
\ref{comb} with each galaxy's profile as a separate color.  The uncertainties
on the points are the $\Delta \chi^2=1$ uncertainties from Figure \ref{all}.
We have scaled each galaxy's profile relative to an arbitrary $r^{-1}$ profile.
In this way the shape of each profile is preserved and only the 
height has been adjusted to reduce the scatter.  We fit a line to the
$\log \rho_{DM}$ profiles and determine that the slope $\alpha = 1.2 \pm 0.5$.
We also restrict our fit to only points in the profile where we have kinematics
(dotted line in Figure \ref{comb}) and find a similar slope of 
$\alpha = 0.9 \pm 0.5$.  

We conclude from Figure \ref{comb} that the \emph{average} dark matter 
profile in the dSphs is similar to an $r^{-1}$ profile.  
However, when we model
each galaxy individually, we find a variety of profiles that are
different from the mean $r^{-1}$ profile.  Our interpretation of this
observation is that variations in their
individual formation histories cause galaxies to scatter from the
average profile.  Only when multiple galaxies are averaged together
does it become clear they follow a combined $r^{-1}$ profile.  This
single power-law profile compares well with the predicted NFW profile
in the inner portion of the plot.  However, at larger radii ($\gsim
1~$kpc in dwarf galaxies) the NFW profile becomes steeper than
$r^{-1}$ \citep{spr08}.  More data are needed at both large and small
radii to further explore this.

\begin{acknowledgements}
K.G. acknowledges support from NSF-0908639.
This work would not be possible without the
state-of-the-art supercomputing facilities at the Texas Advanced Computing
Center (TACC).  We also thank Matt Walker and the MMFS Survey
team for making their radial velocities publicly available.

\end{acknowledgements}

\bibliographystyle{apj}

\begin{thebibliography}{50}
\expandafter\ifx\csname natexlab\endcsname\relax\def\natexlab#1{#1}\fi

\bibitem[{{Arraki} {et~al.}(2012){Arraki}, {Klypin}, {More}, \&
  {Trujillo-Gomez}}]{arr12}
{Arraki}, K.~S., {Klypin}, A., {More}, S., \& {Trujillo-Gomez}, S. 2012, ArXiv
  e-prints, astro-ph/1212.6651

\bibitem[{{Blumenthal} {et~al.}(1986){Blumenthal}, {Faber}, {Flores}, \&
  {Primack}}]{blu86}
{Blumenthal}, G.~R., {Faber}, S.~M., {Flores}, R., \& {Primack}, J.~R. 1986,
  \apj, 301, 27

\bibitem[{{Borriello} \& {Salucci}(2001)}]{bor01}
{Borriello}, A., \& {Salucci}, P. 2001, \mnras, 323, 285

\bibitem[{{Breddels} {et~al.}(2012){Breddels}, {Helmi}, {van den Bosch}, {van
  de Ven}, \& {Battaglia}}]{bre12}
{Breddels}, M.~A., {Helmi}, A., {van den Bosch}, R.~C.~E., {van de Ven}, G., \&
  {Battaglia}, G. 2012, ArXiv e-prints, astro-ph/1205.4712

\bibitem[{{Burkert}(1995)}]{bur95}
{Burkert}, A. 1995, \apjl, 447, L25+

\bibitem[{{Coleman} {et~al.}(2005){Coleman}, {Da Costa}, {Bland-Hawthorn}, \&
  {Freeman}}]{col05}
{Coleman}, M.~G., {Da Costa}, G.~S., {Bland-Hawthorn}, J., \& {Freeman}, K.~C.
  2005, \aj, 129, 1443

\bibitem[{{de Blok} {et~al.}(2001){de Blok}, {McGaugh}, {Bosma}, \&
  {Rubin}}]{deb01}
{de Blok}, W.~J.~G., {McGaugh}, S.~S., {Bosma}, A., \& {Rubin}, V.~C. 2001,
  \apjl, 552, L23

\bibitem[{{Gao} {et~al.}(2008){Gao}, {Navarro}, {Cole}, {Frenk}, {White},
  {Springel}, {Jenkins}, \& {Neto}}]{gao08}
{Gao}, L., {Navarro}, J.~F., {Cole}, S., {et~al.} 2008, \mnras, 387, 536

\bibitem[{{Gebhardt} \& {Thomas}(2009)}]{geb09}
{Gebhardt}, K., \& {Thomas}, J. 2009, \apj, 700, 1690

\bibitem[{{Gebhardt} {et~al.}(1996){Gebhardt}, {Richstone}, {Ajhar}, {Lauer},
  {Byun}, {Kormendy}, {Dressler}, {Faber}, {Grillmair}, \& {Tremaine}}]{geb96}
{Gebhardt}, K., {Richstone}, D., {Ajhar}, E.~A., {et~al.} 1996, \aj, 112, 105

\bibitem[{{Gebhardt} {et~al.}(2000){Gebhardt}, {Bender}, {Bower}, {Dressler},
  {Faber}, {Filippenko}, {Green}, {Grillmair}, {Ho}, {Kormendy}, {Lauer},
  {Magorrian}, {Pinkney}, {Richstone}, \& {Tremaine}}]{geb00}
{Gebhardt}, K., {Bender}, R., {Bower}, G., {et~al.} 2000, \apjl, 539, L13

\bibitem[{{Gebhardt} {et~al.}(2003){Gebhardt}, {Richstone}, {Tremaine},
  {Lauer}, {Bender}, {Bower}, {Dressler}, {Faber}, {Filippenko}, {Green},
  {Grillmair}, {Ho}, {Kormendy}, {Magorrian}, \& {Pinkney}}]{geb03}
{Gebhardt}, K., {Richstone}, D., {Tremaine}, S., {et~al.} 2003, \apj, 583, 92

\bibitem[{{Goerdt} {et~al.}(2006){Goerdt}, {Moore}, {Read}, {Stadel}, \&
  {Zemp}}]{goe06}
{Goerdt}, T., {Moore}, B., {Read}, J.~I., {Stadel}, J., \& {Zemp}, M. 2006,
  \mnras, 368, 1073

\bibitem[{{Irwin} \& {Hatzidimitriou}(1995)}]{irw95}
{Irwin}, M., \& {Hatzidimitriou}, D. 1995, \mnras, 277, 1354

\bibitem[{Jardel \& Gebhardt(2012)}]{jar12}
Jardel, J.~R., \& Gebhardt, K. 2012, \apj, 746, 89

\bibitem[{{Jardel} {et~al.}(2013){Jardel}, {Gebhardt}, {Fabricius}, {Drory}, \&
  {Williams}}]{jar13}
{Jardel}, J.~R., {Gebhardt}, K., {Fabricius}, M.~H., {Drory}, N., \&
  {Williams}, M.~J. 2013, \apj, 763, 91

\bibitem[{{Kleyna} {et~al.}(2002){Kleyna}, {Wilkinson}, {Evans}, {Gilmore}, \&
  {Frayn}}]{kle02}
{Kleyna}, J., {Wilkinson}, M.~I., {Evans}, N.~W., {Gilmore}, G., \& {Frayn}, C.
  2002, \mnras, 330, 792

\bibitem[{{Lee} {et~al.}(2009){Lee}, {Yuk}, {Park}, {Harris}, \&
  {Zaritsky}}]{lee09}
{Lee}, M.~G., {Yuk}, I.-S., {Park}, H.~S., {Harris}, J., \& {Zaritsky}, D.
  2009, \apj, 703, 692

\bibitem[{{Lianou} {et~al.}(2011){Lianou}, {Grebel}, \& {Koch}}]{lia11}
{Lianou}, S., {Grebel}, E.~K., \& {Koch}, A. 2011, \aap, 531, A152

\bibitem[{{Maraston}(2005)}]{mar05}
{Maraston}, C. 2005, \mnras, 362, 799

\bibitem[{{Mateo}(1998)}]{mat98}
{Mateo}, M.~L. 1998, \araa, 36, 435

\bibitem[{{Merritt} {et~al.}(2005){Merritt}, {Navarro}, {Ludlow}, \&
  {Jenkins}}]{mer05}
{Merritt}, D., {Navarro}, J.~F., {Ludlow}, A., \& {Jenkins}, A. 2005, \apjl,
  624, L85

\bibitem[{{Navarro} {et~al.}(1996{\natexlab{a}}){Navarro}, {Eke}, \&
  {Frenk}}]{nav96b}
{Navarro}, J.~F., {Eke}, V.~R., \& {Frenk}, C.~S. 1996{\natexlab{a}}, \mnras,
  283, L72

\bibitem[{{Navarro} {et~al.}(1996{\natexlab{b}}){Navarro}, {Frenk}, \&
  {White}}]{nav96}
{Navarro}, J.~F., {Frenk}, C.~S., \& {White}, S.~D.~M. 1996{\natexlab{b}},
  \apj, 462, 563

\bibitem[{{Navarro} {et~al.}(2010){Navarro}, {Ludlow}, {Springel}, {Wang},
  {Vogelsberger}, {White}, {Jenkins}, {Frenk}, \& {Helmi}}]{nav10}
{Navarro}, J.~F., {Ludlow}, A., {Springel}, V., {et~al.} 2010, \mnras, 402, 21

\bibitem[{{Odenkirchen} {et~al.}(2001){Odenkirchen}, {Grebel}, {Harbeck},
  {Dehnen}, {Rix}, {Newberg}, {Yanny}, {Holtzman}, {Brinkmann}, {Chen},
  {Csabai}, {Hayes}, {Hennessy}, {Hindsley}, {Ivezi{\'c}}, {Kinney},
  {Kleinman}, {Long}, {Lupton}, {Neilsen}, {Nitta}, {Snedden}, \&
  {York}}]{ode01}
{Odenkirchen}, M., {Grebel}, E.~K., {Harbeck}, D., {et~al.} 2001, \aj, 122,
  2538

\bibitem[{{Persic} {et~al.}(1996){Persic}, {Salucci}, \& {Stel}}]{per96}
{Persic}, M., {Salucci}, P., \& {Stel}, F. 1996, \mnras, 281, 27

\bibitem[{{Piatek} {et~al.}(2007){Piatek}, {Pryor}, {Bristow}, {Olszewski},
  {Harris}, {Mateo}, {Minniti}, \& {Tinney}}]{pia07}
{Piatek}, S., {Pryor}, C., {Bristow}, P., {et~al.} 2007, \aj, 133, 818

\bibitem[{{Piatek} {et~al.}(2003){Piatek}, {Pryor}, {Olszewski}, {Harris},
  {Mateo}, {Minniti}, \& {Tinney}}]{pia03}
{Piatek}, S., {Pryor}, C., {Olszewski}, E.~W., {et~al.} 2003, \aj, 126, 2346

\bibitem[{{Pietrzy{\'n}ski} {et~al.}(2008){Pietrzy{\'n}ski}, {Gieren},
  {Szewczyk}, {Walker}, {Rizzi}, {Bresolin}, {Kudritzki}, {Nalewajko}, {Storm},
  {Dall'Ora}, \& {Ivanov}}]{pie08}
{Pietrzy{\'n}ski}, G., {Gieren}, W., {Szewczyk}, O., {et~al.} 2008, \aj, 135,
  1993

\bibitem[{{Rix} {et~al.}(1997){Rix}, {de Zeeuw}, {Cretton}, {van der Marel}, \&
  {Carollo}}]{rix97}
{Rix}, H., {de Zeeuw}, P.~T., {Cretton}, N., {van der Marel}, R.~P., \&
  {Carollo}, C.~M. 1997, \apj, 488, 702

\bibitem[{{Schwarzschild}(1979)}]{sch79}
{Schwarzschild}, M. 1979, \apj, 232, 236

\bibitem[{{Sersic}(1968)}]{ser68}
{Sersic}, J.~L. 1968, {Atlas de Galaxias Australes (Cordoba, Argentina:
  Observatorio Astronomico, Univ. Cordoba)}

\bibitem[{{Silverman}(1986)}]{sil86}
{Silverman}, B.~W. 1986, {Density estimation for statistics and data analysis},
  ed. {Silverman, B.~W.}

\bibitem[{{Simon} {et~al.}(2005){Simon}, {Bolatto}, {Leroy}, {Blitz}, \&
  {Gates}}]{sim05}
{Simon}, J.~D., {Bolatto}, A.~D., {Leroy}, A., {Blitz}, L., \& {Gates}, E.~L.
  2005, \apj, 621, 757

\bibitem[{{Simon} \& {Geha}(2007)}]{sim07}
{Simon}, J.~D., \& {Geha}, M. 2007, \apj, 670, 313

\bibitem[{{Siopis} {et~al.}(2009){Siopis}, {Gebhardt}, {Lauer}, {Kormendy},
  {Pinkney}, {Richstone}, {Faber}, {Tremaine}, {Aller}, {Bender}, {Bower},
  {Dressler}, {Filippenko}, {Green}, {Ho}, \& {Magorrian}}]{sio09}
{Siopis}, C., {Gebhardt}, K., {Lauer}, T.~R., {et~al.} 2009, \apj, 693, 946

\bibitem[{{Springel} {et~al.}(2008){Springel}, {Wang}, {Vogelsberger},
  {Ludlow}, {Jenkins}, {Helmi}, {Navarro}, {Frenk}, \& {White}}]{spr08}
{Springel}, V., {Wang}, J., {Vogelsberger}, M., {et~al.} 2008, \mnras, 391,
  1685

\bibitem[{{Tammann} {et~al.}(2008){Tammann}, {Sandage}, \& {Reindl}}]{tam08}
{Tammann}, G.~A., {Sandage}, A., \& {Reindl}, B. 2008, \apj, 679, 52

\bibitem[{{Thomas} {et~al.}(2007){Thomas}, {Jesseit}, {Naab}, {Saglia},
  {Burkert}, \& {Bender}}]{tho07b}
{Thomas}, J., {Jesseit}, R., {Naab}, T., {et~al.} 2007, \mnras, 381, 1672

\bibitem[{{Thomas} {et~al.}(2005){Thomas}, {Saglia}, {Bender}, {Thomas},
  {Gebhardt}, {Magorrian}, {Corsini}, \& {Wegner}}]{tho05}
{Thomas}, J., {Saglia}, R.~P., {Bender}, R., {et~al.} 2005, \mnras, 360, 1355

\bibitem[{{Thomas} {et~al.}(2004){Thomas}, {Saglia}, {Bender}, {Thomas},
  {Gebhardt}, {Magorrian}, \& {Richstone}}]{tho04}
---. 2004, \mnras, 353, 391

\bibitem[{{Valluri} {et~al.}(2004){Valluri}, {Merritt}, \& {Emsellem}}]{val04}
{Valluri}, M., {Merritt}, D., \& {Emsellem}, E. 2004, \apj, 602, 66

\bibitem[{{van den Bosch} {et~al.}(2008){van den Bosch}, {van de Ven},
  {Verolme}, {Cappellari}, \& {de Zeeuw}}]{vdb08}
{van den Bosch}, R.~C.~E., {van de Ven}, G., {Verolme}, E.~K., {Cappellari},
  M., \& {de Zeeuw}, P.~T. 2008, \mnras, 385, 647

\bibitem[{{van der Marel} {et~al.}(1998){van der Marel}, {Cretton}, {de Zeeuw},
  \& {Rix}}]{vdm98}
{van der Marel}, R.~P., {Cretton}, N., {de Zeeuw}, P.~T., \& {Rix}, H. 1998,
  \apj, 493, 613

\bibitem[{{Walcher} {et~al.}(2003){Walcher}, {Fried}, {Burkert}, \&
  {Klessen}}]{wal03}
{Walcher}, C.~J., {Fried}, J.~W., {Burkert}, A., \& {Klessen}, R.~S. 2003,
  \aap, 406, 847

\bibitem[{{Walker} {et~al.}(2008){Walker}, {Mateo}, \& {Olszewski}}]{wal08}
{Walker}, M.~G., {Mateo}, M., \& {Olszewski}, E.~W. 2008, \apjl, 688, L75

\bibitem[{{Walker} {et~al.}(2009{\natexlab{a}}){Walker}, {Mateo}, \&
  {Olszewski}}]{wal09}
---. 2009{\natexlab{a}}, \aj, 137, 3100

\bibitem[{{Walker} {et~al.}(2009{\natexlab{b}}){Walker}, {Mateo}, {Olszewski},
  {Pe{\~n}arrubia}, {Wyn Evans}, \& {Gilmore}}]{wal09b}
{Walker}, M.~G., {Mateo}, M., {Olszewski}, E.~W., {et~al.} 2009{\natexlab{b}},
  \apj, 704, 1274

\bibitem[{{Walker} \& {Pe{\~n}arrubia}(2011)}]{wal11}
{Walker}, M.~G., \& {Pe{\~n}arrubia}, J. 2011, \apj, 742, 20

\end{thebibliography}

\end{document}